\documentstyle[aps,prb,multicol]{revtex}
\begin{document}

\title{Spectral flow in vortex dynamics of d-wave superconductors}
\author{Yu.G.Makhlin}
\address{Department of Physics\\
University of Illinois at Urbana-Champaign\\
1110, West Green St, Urbana, IL, 61801\\
and\\
L.D. Landau Institute for Theoretical Physics, \\
Kosygin Str. 2, 117940 Moscow, Russia
}

\date{June 18, 1997}
\maketitle

\begin{abstract}
We calculate Ohmic and Hall flux flow conductivities in the mixed state
of d-wave superconductors taking into account the spectral flow along
the anomalous branch of the states bound to vortex cores. Two
parameters governing the strength of the spectral flow at low
temperatures are found. They are determined by the ratios of the width
of the localized levels $1/\tau$ and 1) the interlevel spacing in the
core, and 2) $\Delta_0^2/E_F$. These parameters determine three
regimes of vortex dynamics.
In the moderately clean limit the contribution of the
spectral flow to the conductivities coincides with that for
s-superconductors. In the superclean regime in low magnetic fields both
Ohmic and Hall conductivities acquire $\tau$-independent universal
values. The contribution of the spectral flow in the superclean regime
is suppressed in higher fields. We also discuss the case of higher
temperatures.
\end{abstract}


\begin{multicols}{2}

\section{Introduction}

The problem of vortex motion in superconductors/superfluids and
flux-flow Hall effect attracted attention of researchers for years.
The renewed interest to this problem is due to recent developments in
the physics of the Hall effect in high-temperature superconductors and
in the investigations of the superfluid $^3$He.  A microscopic theory
of mutual friction (friction between the superfluid and normal
components in presence of quantized vortices) was developed long time
ago.\cite{Kopnin} This theory (see also
Refs.\onlinecite{KopninLopatin,VolovikSpFlowOrig,Stone,OFGB}) enables us to
calculate the forces on a vortex (or, equivalently, Ohmic and Hall
conductivities) as functions of microscopic and external parameters.
The temperature dependence of the forces on a vortex predicted
theoretically \cite{Parts} is in good agreement with the experimental data
\cite{HallHook} for $^3$He-B. The theory also proposes an explanation
for the sign reversal of the Hall angle as an intrinsic property of the
vortex flow in conventional and high-$T_c$
superconductors.\cite{KopninLopatin,OFGB}
This theory applies to the regime of free flux flow when pinning is not
important. In superconductors this region can be reached experimentally
either by preparation of high purity untwinned samples or by using high
density driving currents.

However, all the previous calculations were performed for the isotropic
s-pairing.  Since this theory is used for analysis of the data for
high-$T_c$ materials it is necessary to understand whether its
arguments are sensitive to the symmetry of the pairing state and the
presence of gap nodes.

The total non-dissipative force on a vortex
can be subdivided into the following three contributions of different
physical origin (see Ref.\onlinecite{3forces} and references therein):
\begin{eqnarray}
{\bf F}=&&\rho{\vec\kappa}\times({\bf v}_L-{\bf v}_s)+
\rho_n{\vec\kappa}\times({\bf v}_s-{\bf v}_n)\nonumber\\
&&+C{\vec\kappa}\times({\bf v}_n-{\bf v}_L).
\label{eq1}
\end{eqnarray}
where $\rho$, $\rho_n$ are the densities of the fluid and the normal
component, respectively, $\vec\kappa$ is the vector of the circulation
of the superfluid velocity around the vortex directed along the vortex
axis, $\kappa=\pi\hbar/m$ for a 1-quantum vortex.  ${\bf v}_n$ and
${\bf v}_s$ are the velocities of the normal and superfluid components
far from the vortex core and ${\bf v}_L$ is the velocity of the vortex
(the vortex lattice).  The first term in (\ref{eq1}) is the Magnus
force describing the transfer of momentum from the superfluid
vacuum to the vortex. The second one is the Iordanskii force
\cite{Iordanskii} due to asymmetric scattering of excitations by the
vortex core. These first two terms could be rewritten as
${\vec\kappa}\times [\rho_s({\bf v}_s-{\bf v}_L)+\rho_n({\bf v}_n-{\bf
v}_L)]$ and thought about as contributions of the superfluid and normal
components.

In this article we concentrate on the last term in (\ref{eq1}).  This
term describes the transfer of momentum between the heat bath and the
bound states in the vortex core specific for Fermi superfluids,
\cite{Caroli} which is due to the spectral flow (SF) along the
anomalous branch of bound states.\cite{VolovikSpFlowOrig}  For s-wave
superconductors the coefficient $C$ in front of this term depends on
the ratio of the interlevel distance \cite{Caroli} in the core
$\omega_0\sim \Delta^2/E_F$ to the inverse relaxation time $\tau$ for
the bound states. When the width of the levels $1/\tau$ is greater than
the interlevel spacing $\omega_0$, the spectrum is continuous and the
SF is allowed.\cite{VolovikSpFlowOrig} In this limit $C=C_0$, which is
the density of the fluid in the normal state ($C_0=mk_F^2/2\pi$ in 2D,
$C_0=mk_F^3/3\pi^2$ in 3D). The difference
$\rho-C_0\sim\rho(\Delta/E_F)^2$ is small and ${\bf v}_L$ dependence of
the total force (\ref{eq1}) is weak.  In the opposite limit
$\omega_0\tau\gg1$ the spectrum is discrete and the SF \cite{SF-term}
is suppressed, $C\to 0$.

Another explanation of the dependence of the strength of the spectral
flow on $\tau$ was proposed by Stone.\cite{Stone} He showed that
quasiclassical bound states inside the vortex core rotate with the
angular velocity $\omega_0$ and this rotation suppresses the spectral
flow. However, relaxation effectively prevents the rotation if the
relaxation time $\tau$ is less than the period of rotation
$2\pi/\omega_0$. Therefore, the SF is suppressed in the limit
of large $\omega_0\tau$ and allowed in the opposite limit.
In the regime of free flux flow the only contribution to the
dissipative force is also due to the bound states. The dissipative
force is small in both limits and is maximal at $\omega_0\tau\sim1$.

At finite $T\sim T_c$ a contribution of the continuous spectrum of
states above the gap to $C$ is also essential. Since the interlevel
distance above the gap is given by the cyclotron frequency
$\omega_c=|e|H/m^*c\sim\omega_0(H/H_{c2})$, this contribution of the
continuous \cite{about_cont} spectrum is suppressed \cite{Parts} in the limit
$\omega_c\tau\gg1$.

In this article we address the question of applicability of the theory
to the d-wave case,\cite{VolPoisson} namely,
to $d_{x^2-y^2}$-pairing state,
which is believed to represent the order parameter in high-$T_c$
materials, \cite{dwave} though the analysis for an arbitrary pairing
state with points of gap nodes in 2D (or lines in 3D) would be
similar. The structure of the states bound to vortex cores in d-superconductors
is different from that in s-wave materials.\cite{Volovik93} In
particular, low-energy states are extended in directions of gap nodes.
In addition, at energies below the maximum value of the gap $\Delta_0$
bound states coexist with the states in the continuous spectrum.
So, one can expect new features in the contribution of these low-lying levels
to vortex dynamics.

The interlevel spacing in cores of d-wave vortices \cite{Volovik93}
\begin{equation}
\epsilon_0\simeq\frac{2\Delta'^2}{v_Fk_F}\sqrt{\frac{H}{H_{c2}}}
\ln\left(\frac{H}{H_{c2}}\right)
\label{eq2} 
\end{equation}
is much  less than in s-superconductors:
$\omega_c\ll\epsilon_0\ll\Delta_0^2/E_F$ if $H\ll H_{c2}$.
Comparison of $\epsilon_0$ to the width $1/\tau$ of the levels suggests
the dimensionless parameter $\epsilon_0\tau$ governing the strength of
the SF.

On the other hand, the angular velocity of rotation of Andreev bound
states $\omega_0(\theta)$ is angle-dependent in d-wave superconductors
(it follows the angle dependence of the gap; see Section 2). Its
maximal value $\omega_0^{max}$ on the order of $\Delta_0^2/E_F$
($\Delta_0$ is the maximal value of the gap) is much greater than
$\epsilon_0$. So, this approach proposes a different scale for $1/\tau$
as a boundary between the regimes of extreme and suppressed SF. In
addition, near the gap nodes the interlevel distance $\omega_0(\theta)$
is very small (see Section 2), i.e., the density of states is
large.\cite{Volovik93} So the contribution of this region needs special
attention.

One of our purposes is to understand, which parameter governs the
strength of the SF in the d-wave case. In particular, this question is
important for analysis of experimental data: one should locate the
boundary between the regimes of extreme and suppressed SF.  In
addition, it is interesting to estimate the force on a vortex and Ohmic
and Hall conductivities in various limiting cases to compare the
results to experiments and previous theories. In particular, we look
for results different from those for s-superconductors. We found new
regimes of vortex dynamics in the superclean limit in superconductors
with gap nodes (d-wave), different from those in isotropic (s-wave)
superconductors. The SF force depends on both parameters discussed
above, $\epsilon_0\tau$ and $\Delta_0^2\tau/E_F$ (as well as on
$\omega_c\tau$ at finite $T$).

It turns out that it is convenient to explore this problem using the
kinetic equation approach developed recently by Stone \cite{Stone} for
s-wave vortices. We apply it to the d-wave case and find
the solution of the equation. To use the kinetic equation (KE) one should
first investigate the spectrum of the bound states (Section 2). Then, in
Section 3, we solve the KE and discuss its implications on
the force in various limiting cases in Section 4.
We consider the 2D case (applicable directly to thin films and pancake
vortices). Generalization to 3D is straightforward (see discussion).

In Section 5 we explore the KE in terms of the density matrix in the
basis of exact eigenstates of Bogolyubov-de Gennes (BdG) hamiltonian,
and evaluate parameters governing the strength of the SF in a different
way. Finally, we summarize and discuss our results in Section 6.

\section{Spectrum of the bound states}

In this Section we describe states inside and near vortices
in d-wave superconductors. We consider a material with a circular
particle-like Fermi surface (generalization to the case of a hole-like
Fermi surface or coexisting hole- and particle-like parts of a Fermi
surface is straightforward).
We assume that the coherence length $\xi_0$
in the superconducting state is much greater than the inverse Fermi
momentum $k_F^{-1}$ (or, equivalently, $\Delta_0\ll E_F$). Even in
high-$T_c$ materials, where $E_F/\Delta_0$ is not very large, it is still of
the order of 7--10. Therefore we can use the quasiclassical approach and
solve the BdG equations in Andreev approximation. We also assume that
the magnetic field $H\ll H_{c2}$ so that vortices are far apart (and, of
course, $H>H_{c1}$ so that the sample is in the mixed state).

In $d_{x^2-y^2}$-wave state the distribution of the order parameter in 
the vicinity of a vortex is given by
\begin{equation} 
\Delta({\bf r},\theta)=\Delta_0(r)e^{i\varphi}(\hat k^2_a -\hat k^2_b)
\end{equation}
where $\hat a$ and $\hat b$ are the crystal axes,
$r$, $\varphi$ are polar coordinates in the position space. We
put $\hat a=\hat x$, $\hat b=\hat y$. The absolute
value of the gap $\Delta_0(r)$ equals zero at the center of the core and
saturates at the bulk value $\Delta_0$ far from the core, at distances 
$r\gg\xi_0$. In the bulk
$\Delta(\theta)=\Delta_0\cos(2\theta)$, where the angle $\theta$ is
defined by ${\bf k}=k_F(\cos\theta;\sin\theta)$, 
and there are 4 gap nodes (in 2D
case). Close to a gap node $\theta_0$ the order parameter is given by
$\Delta(\theta)=\Delta'(\theta-\theta_0)$. 

We consider clean superconductors ($\Delta(\theta)\tau(\theta)\gg1$) so
that the motion of quasiparticles inside and near the core is ballistic.
In the quasiclassical approach the states are described by the direction of
the momentum $\theta$ and the
spatial position of the line, on which the state is situated. This spatial 
position can be described by the impact parameter $b$
or, equivalently, by the angular momentum $l=k_Fb$.
In d-wave vortices bound states could extend far outside vortex
cores.\cite{Volovik93}
The characteristic
extension of these states in the direction of their momentum
is of the order of the angle-dependent coherence length 
$\xi(\theta)=v_F/|\Delta(\theta)|$ where $\Delta(\theta)$ is the
angle-dependent d-wave superconducting order parameter.

Below the gap there is one chiral branch of bound states.  Due to a
special symmetry of (1D) BdG hamiltonian in Andreev approximation for a
state situated at a diameter ($l=0$) the eigenenergy of the bound state
is exactly zero,\cite{Volovik93} and the eigenfunction is
\begin{equation}
\Psi(x)\propto \left({1\atop i}\right)
\exp\left(-\frac{1}{v_F}\int\limits_0^x dy
\Delta(y,\theta)\right)
\end{equation}
with the extension of the order of $\xi(\theta)$.
In Ref.\onlinecite{Volovik93}
the spectrum of states with impact parameters smaller than the
vortex core size was calculated perturbatively, the correction to the
energy being
\begin{equation}
E=-b\frac{ 2\int dx \frac{\Delta(x,\theta)}{x}
\exp\left(-\frac{2}{v_F}\int\limits_0^x dy\Delta(y,\theta)\right) }
{\int dx\exp\left(-\frac{2}{v_F}\int\limits_0^x dy\Delta(y,\theta)\right) } 
. \label{disp}
\end{equation}
For small values of the impact parameter the logarithmically divergent
integral in the
numerator of the previous expression should be cut off from below at the
core size $\xi_0$, that is,
\begin{equation} E(l,\theta)=-\omega_0(\theta,0) l ,
\label{SpectrumSmallEn}
\end{equation}
where the interlevel distance is \cite{Volovik93}
\begin{equation}
\omega_0(\theta,0)=\Omega_0
(\theta-\theta_0)^2\ln(1/|\theta-\theta_0|)
\label{omegasmall}
\end{equation}
in a vicinity of a gap node $\theta_0$, $\Omega_0=2\Delta'^2/v_Fk_F$.
Far from the nodes $\omega_0(\theta)$ is of the order of $\Omega_0$.
However, for impact parameters larger than the core size the lower
cutoff is given by the impact parameter itself. Therefore, to the
logarithmic accuracy one finds that
\begin{eqnarray}
E(l,\theta)\sim -\Omega_0 (\theta-\theta_0)^2
\ln\left( \frac{\xi(\theta)}{\hbox{\rm max}(\xi_0,l/k_F)} \right)
l ,\label{SpectrumLargeEn}\\
\qquad \xi_0\ll b\ll\xi(\theta) .\nonumber
\end{eqnarray}
At large values of the impact parameter $|b|\gg\xi(\theta)$ the energy
of the bound state saturates at the gap value $\Delta(\theta)$, the
asymptotic behavior being given by:\cite{Rainer}
\begin{equation}
E(l,\theta)\sim -|\Delta(\theta)|\mbox{sign}(l)+ const\frac{E_F}{l} ,
\label{Easympt}
\end{equation}
where $const$ is of the order of unity.
Saturation of the spectrum occurs at $l\sim
E_F/|\Delta(\theta)|\sim k_F\xi(\theta)$ where spatial variations of
the order parameter along the quasiclassical trajectory are slow.

In (charged) superconductors the effect of magnetic field on motion of
quasiparticles leads to an additional term in the dispersion relation
(\ref{disp}): $\delta E=-\omega_cl$. This term is important for
excitations with momenta close to gap nodes, where $\omega_0(\theta,0)$
is small.

Close to the nodes ($\hat{\bf k}=\pm\hat a,\pm\hat b$) the
extension $\xi(\theta)$ of the wave function of a localized state is
very large and can exceed the intervortex distance $R_v$, which
establishes the characteristic distance of the changes in the order
parameter and superfluid velocity in the vortex lattice.
For angles $\theta$ close to a
gap node $\theta_0$ such that $|\theta-\theta_0|<\theta_H$, where
$\theta_H=\Delta'/(v_FR_v)\sim\xi_0/R_v \sim\sqrt{H/H_{c2}}$,
corresponding states are not localized any more and should be treated
separately.  For $|\theta-\theta_0|\sim\theta_H$ the value
$\omega_0(\theta)$ becomes comparable to the cyclotron frequency
$\omega_c$ of the external field, which also contributes to
$\omega_0(\theta)$. We model this region by putting
$\omega_0(\theta)\equiv\omega_0(\theta_H)\sim\omega_c$
for these angles close to gap nodes.
The only characteristic of the spectrum in this
region, which affects the value of the force, is the interlevel distance
$\epsilon_0$ (see below).

In addition to the just described branch of bound states there exist
branches of the states with energies above the gap $\Delta(\theta)$,
that is, in the continuous spectrum. For a one-quantum vortex there is
one chiral branch in the continuous spectrum, which crosses all
energy levels once.\cite{Kopnin1} The interlevel spacing in the
continuous spectrum is $\omega_0=\omega_c$.

For later analysis we need to know the angle dependence of the
interlevel spacing
\begin{equation}
\omega_0(\theta,E)=-\left(\frac{\partial E}{\partial l}\right)_\theta
\end{equation}
for different energies $E$.  For very small energies $|E|\ll
\Delta'\theta_H$ this dependence is described by
Eq.(\ref{omegasmall}) above with the just discussed condition of
saturation near nodes. For higher energies $\Delta'\theta_H\ll |E| <
\Delta_0$ the spectrum near the gap nodes is ``continuous'':
\begin{equation}
\omega_0(\theta,E)=\omega_c, \qquad |\theta-\theta_0|<\theta_E
\label{SpBetween}
\end{equation}
(where $\theta_E$ is defined by $|\Delta(\theta_E)|=|E|$,
$\theta_E<\pi/4$; $\theta_E=|E|/\Delta'$ for $|E|\ll\Delta_0$). On
approaching this region from outside $\omega_0(\theta,E)$ decreases.
To investigate the behavior of $\omega_0(\theta,E)$ as we approach this
vicinity from outside we use the expression (\ref{Easympt}) and find
the relationship between $l$ and $\theta$ at given $E$ as
$\theta\to\theta_E$. We find that
$l\propto(\theta-\theta_E)^{-1}\to\infty$ and
$\omega_0\sim\Omega_0(\theta-\theta_E)^2$.  However, for $l>k_FR_v$ the
states are sensitive to the velocity fields of other vortices in the
lattice, and therefore $\omega_0(\theta,E)$ saturates at a finite
value. This value is on the order of $\omega_c$, which also contributes
to $\partial E/\partial l$ (\ref{SpBetween}). So, the behavior of
$\omega_0(\theta,E)$ as $\theta$ approaches $\theta_E$ from outside is
quantitatively similar to the behavior of $\omega_0(\theta,0)$ as
$\theta\to 0$. 

As $|E|\to\Delta_0$ the region of the continuous spectrum
(\ref{SpBetween}) grows and at $|E|>\Delta_0$ the interlevel distance
$\omega_0(\theta,E)=\omega_c$ for all $\theta$.  The difference in the
behavior of $\omega_0(\theta,E)$ and, in particular, coexistence of
the regions of continuous and discrete spectrum at the same energy $E$,
determines difference in the solutions of the KE at different energies
(and therefore different temperatures).

We can use the quasiclassical approximation as long as the temperature is
larger than the interlevel spacing. The maximal value of
$\omega_0(\theta,E)$ is on the order of $\Omega_0$, i.e., we need
$T\gg\Omega_0$.

Now, as we know the structure of the spectrum we can proceed to
consideration of the dynamics of the distribution function.

\section{Kinetic equation for the distribution function} 

We consider the problem in the frame of the vortex lattice, so
that ${\bf v}_L=0$.
To calculate the effect of the SF (linear in the
velocity of the normal component with respect to the vortex lattice
${\bf v}_n-{\bf v}_L$) we put the external superflow ${\bf v}_s=0$
but keep the velocity of the normal component ${\bf v}_n$ finite.
To the first order, the 
SF force is a linear function of ${\bf v}_n$, posessing 
$D_4$-symmetry of the pairing state. Therefore, the force on a vortex
is determined by two parameters:
\begin{equation}
{\bf F}_{SF}=d^\perp_{SF}C_0{\vec \kappa}\times {\bf v}_n +
d^\parallel_{SF}C_0\kappa{\bf v}_n .
\end{equation}
In particular, the longitudinal and transverse components of the force
do not depend on the direction of ${\bf v}_n$. For simplicity we
choose ${\bf v}_n=v_n\hat x$.
The dynamics of the distribution function (or the density matrix)
can be described by the KE.
Since $\theta$ and $l$ are canonically conjugate,
this equation reads (cf. Ref.\onlinecite{Stone}):
\begin{equation}
\frac{\partial n(l,\theta)}{\partial t}
-\frac{\partial E}{\partial\theta}\frac{\partial n}{\partial l}
+\frac{\partial E}{\partial l}\frac{\partial n}{\partial\theta}
=-\frac{n-n^{eq}}{\tau} ,
\label{KE0}
\end{equation}
where $n^{eq}=n_F(E-{\bf v}_n{\bf k})=n_F(E(l,\theta)-v_nk_F\cos\theta)$ 
is the distribution function in equilibrium with the heat bath (the normal
component), $n_F$ is the Fermi distribution function.
The treatment of the collisions in the relaxation time approximation is,
of course, not accurate. Nevertheless, we believe that it qualitatively
correctly describes the physics of the system.  The relaxation time
$\tau$ can be angle and energy-dependent, and we can easily incorporate
this dependence into our results of this Section.  However, for
investigation of limiting cases we put $\tau(\theta)=const$.

We study the stationary case and put $\partial_tn=0$ in
Eq.(\ref{KE0}). If we change the variables $l,\theta$ to $E,\theta$,
the equation takes the form
\begin{equation}
\frac{\partial
n(E,\theta)}{\partial\theta}=\frac{n(E,\theta)-n^{eq}(E,\theta)}
{\omega_0(E,\theta)\tau} . \label{KE}
\end{equation}
The equation (\ref{KE}) can be easily solved:
\begin{eqnarray}
n(E,\theta)=C(E)&&\exp\left(\int\limits_0^\theta \frac{d\varphi}
{\omega_0(E,\varphi)\tau}\right)-
\label{nsolution}  \\
\int\limits_0^\theta\frac{d\chi}{\omega_0(E,\chi)\tau}
n_F(E&&-v_nk_F\cos\chi)\exp\left(\int\limits_\chi^\theta
\frac{d\varphi}{\omega_0(E,\varphi)\tau}\right) .\nonumber
\end{eqnarray}
The energy-dependent constant $C(E)$ should be determined from the
condition of $2\pi$-periodicity of $n(E,\theta)$ in $\theta$.

Since we
know the distribution function we can substitute it into the expression
for the SF force (the force on vortex from the heat bath,
which in the case of a crystal is the lattice with the impurities).
Since the momentum of the excitations is given by:\cite{Stone}
\[
{\bf P}=\frac{1}{2}\int\frac{d\theta dl}{2\pi}k_F(\cos\theta;\sin\theta)
(n-n_0)
\]
[where $n_0(l,\theta)=\Theta(l)$ stands for normalization purposes and ensures
that ${\bf P}=0$ in equilibrium at $T=0$;
$\Theta(l)=0$ for $l<0$, $\Theta(l)=1$ for $l>0$; the prefactor $1/2$
compensates for the double counting of particles and holes],
the contribution of the heat
bath to the force \cite{swaveresult} is [cf. (\ref{KE0})]
\begin{equation}
{\bf F}_{SF}=\frac{1}{2}\int\frac{d\theta dl}{2\pi}k_F
(\cos\theta;\sin\theta)\frac{n^{eq}-n}{\tau}.
\label{result}
\end{equation}
To take into account all the branches we should sum up the expressions
(\ref{result}) for all of them. However, non-chiral branches never
contribute to the force.

\section{Reactive and dissipative forces in various limits}

In principle one can substitute the solution (\ref{nsolution}) of the 
KE into the expression for the force (\ref{result}), and if the
temperature dependence of the gap and the relaxation time is known, one
obtains the temperature dependence of the SF force. However,
it is interesting to investigate the behavior of the force in different
physical limits analytically. 

We are looking for an expression for the force to the first order in
$v_n$, so we expand the KE to this order. Namely, we characterize the
deviation of the distribution function from its equilibrium value at
$v_n=0$ by a function $\nu$:
\begin{equation}
n(E,\theta)=n_F(E)-\frac{\partial n_F}{\partial E} v_nk_F \nu(E,\theta) .
\end{equation}
The stationary KE takes the form
\begin{equation}
\frac{\partial
\nu}{\partial\theta}=\frac{\nu-\nu^{eq}}{\omega_0(E,\theta)\tau}
\label{KEnu0}
\end{equation}
where $\nu^{eq}=\cos\theta$. This equation has a unique solution, which
is an odd function of $\theta$:
$\nu(\theta+\pi)=-\nu(\theta)$. The quasiclassical equations of motion
\begin{equation}
\dot\theta=\frac{\partial E}{\partial l},\qquad
\dot l=-\frac{\partial E}{\partial \theta}
\label{quas_motion}
\end{equation}
describe the motion along a curve of fixed energy in
$(\theta,l)$-plane, and the KE (\ref{KEnu0}) keeps track of variations
of the distribution function $\nu$ along this curve in the stationary
case.  The time of motion along this trajectory is given
by:\cite{t-not}
\begin{equation}
t(\theta)=-\int^\theta\frac{d\varphi}{\omega_0(E,\varphi)}
\label{toftheta}
\end{equation}
and the KE describes the relaxation of $\nu$ towards its equilibrium
($t$-dependent) value:
\begin{equation}
\frac{\partial \nu}{\partial t}=-\frac{\nu-\nu^{eq}}{\tau} .
\label{KEnu}
\end{equation}
The solution of this equation depends on the function $\nu^{eq}(t)$
and the relaxation time $\tau$.

The total time of motion along the closed trajectory 
\[
t_0(E)=\int_0^{2\pi}\frac{d\varphi}{\omega_0(E,\varphi)}
\]
determines the
structure of the energy levels. The quantization of the
quasiclassical motion \cite{Volovik93,KopVol} shows that
in a vicinity of energy $E$ the spectrum is
equidistant with the interlevel distance
\begin{equation}
\epsilon_0(E)= \frac{2\pi}{t_0(E)} .
\end{equation}
For small $|E|\ll \Delta'\theta_H$ the interlevel spacing equals
$\epsilon_0=\epsilon_0(E=0)$ and is given by Eq.(\ref{eq2}).

The solution (\ref{nsolution}) can be simplified in various limits.
The behavior of the force is different in three regimes:  moderately
clean with $\omega^{max}_0\tau\ll 1$, ``very clean'' with
$\epsilon_0\tau\ll1\ll\Omega_0\tau$ and ``extremely clean'' when
$\epsilon_0\tau\gg1$. Usually the term ``superclean'' is used for the
regime with $\Omega_0\tau\gg1$. However in the mixed state of d-wave
materials vortex dynamics is different in two subregimes. Therefore we
use the terms ``very clean'' and ``extremely clean'' for these two
subregimes.  Note that $\epsilon_0\tau$ depends on external field $H$
and temperature, while $\Omega_0\tau$ is determined by temperature only.
In subsections A, B, C below we consider these three regimes at low
temperatures $T\ll T_c\theta_H$ [this condition is consistent with
$T\gg\Omega_0$ if $H/H_{c2}\gg(\Delta_0/E_F)^2$].  At corresponding
energies $|E|\ll\Delta'\theta_H$ the quasiclassical interlevel spacing
is given by $\omega_0(\theta,0)$.  The case of higher $T$ is discussed
in subsection D.

\subsection{Moderately clean limit}

In the moderately clean case 
the parameter $\omega_0\tau$ is small for all values of $\theta$:
$\omega_0^{max}\tau\ll1$, and $\nu^{eq}(t)$ is a slow function of $t$
(on the scale of $\tau$). Therefore,
the solution $\nu$ of the KE everywhere is very close to $\nu^{eq}$.
Simple calculation shows that the value of the
SF force in this limit coincides with that in the s-wave
case. The contribution of the bound states is given by:
\begin{eqnarray}
{\bf F}^{\mbox{bound}}_{SF}&\simeq&-\frac{1}{2}\int\frac{d\theta dl}{2\pi}
k_F(\cos\theta;\sin\theta)\frac{\partial n^{eq}}{\partial\theta}
\omega_0(\theta,0)\nonumber\\
&=&2\kappa C_0v_n\int\frac{dEd\theta}{2\pi}
\left(-\frac{\partial n_F}{\partial E}\right)
(\cos\theta;\sin\theta)\sin\theta \nonumber\\
&=&2\kappa C_0v_n
\int\frac{d\theta}{2\pi}(\cos\theta;\sin\theta)\sin\theta\nonumber\\
&&\times[n_F(-|\Delta(\theta)|)-n_F(|\Delta(\theta)|)] .
\end{eqnarray}
In the continuous spectrum only the chiral branch contributes:
\begin{eqnarray}
{\bf F}^{\mbox{cont}}_{SF}&=&4\kappa C_0v_n
\int\frac{d\theta}{2\pi}(\cos\theta;\sin\theta)\sin\theta \nonumber\\
&&\times[1-n_F(-|\Delta(\theta)|)] . 
\end{eqnarray}
So, the total force is
\begin{equation}
{\bf F}_{SF}=C_0{\vec\kappa}\times{\bf v}_n ,
\qquad d^\perp_{SF}=1 .
\label{Ftotal}
\end{equation}

The first correction (in small $\tau$) gives the dissipative force:
\begin{equation}
d^\parallel_{SF}=\tau \langle\omega_0(\theta,0)\rangle ,
\label{MCFdiss}
\end{equation}
where the angle brackets denote averaging over angles $\theta$.

In this limit the SF force almost exactly cancels the Magnus
force, therefore the total non-dissipative force (and the Hall 
conductivity) is given by the force, proportional to the difference
$\rho-C_0$ (see Introduction) and a correction to (\ref{Ftotal}):
\begin{equation}
\delta d^\perp_{SF}=-\tau^2 \langle\omega^2_0(\theta,0)\rangle .
\label{MCdeltaFnd}
\end{equation}
So, the total dissipative and non-disspative forces are of the same
order as in an s-wave superconductor with the same value of the gap
$\Delta_0$.

The result (\ref{Ftotal}) is independent of the structure of the vortex
core. However, the corrections (\ref{MCFdiss}), (\ref{MCdeltaFnd}) depend
on this structure.

\subsection{Very clean limit}

In this limit $\epsilon_0\tau\ll1\ll\omega_0^{max}\tau$ the value of
$\omega_0\tau$ is small near the nodes but it is large far from the
nodes. $t(\theta)$ is almost constant far from the nodes, but it varies
rapidly (on $\tau$ time scale) in their $\theta_\tau$-vicinities where
$\theta_\tau=(\Omega_0\tau)^{-1}$. In an internode region
$\theta_0+a<\theta<\theta_0+\pi/2-a$ (where $\theta_\tau\ll a\ll1$ and
$\theta_0=0,\pi/2,\pi,3\pi/2$ is a gap node) the point in
$(\theta,l)$-plane, describing the quasiclassical motion
(\ref{quas_motion}), spends an amount of time small compared to the
relaxation time $\tau$. It follows from the KE that $\nu$ is almost
constant in such a region (variations of $\nu$ are much less than
unity; we assume, and this assumption is justified by the form of the
solution below, that everywhere on on $\theta$-circle $\nu$ is of the
order of $1$ or less). On the contrary, as the point in
$(\theta;l)$-plane approaches a node its motion gets slower and $\nu$
adjusts to a local value of $\nu^{eq}$ (the positive $t$-direction is
the negative $\theta$-direction, therefore the jump in the value of
$\nu$ occurs on the right-hand side of a node, see below). In a small
$\theta$-vicinity of a gap node $\nu^{eq}$ is almost constant and the
solution of the KE (\ref{KEnu}) can be represented \cite{Exp-strict} as
$\nu=A+Be^{-t/\tau}$ with constant $A$ and $B$. In the very clean
limit, the change in the exponent across a gap node
$t_0/4\tau=\pi/2\epsilon_0\tau$ is very large, and matching the
solutions in two different regions leads to the conclusion that in the
internode region $[\theta_0;\theta_0+\pi/2]$ one has
$\nu=\nu^{eq}(\theta_0+\pi/2)$, and $\nu$ jumps between two constant
values near the nodes. In terms of $\nu$ the force is given by
\begin{eqnarray}
{\bf F}_{SF}=2\kappa C_0v_n&&\int
\frac{dEd\theta}{2\pi}\left(-\frac{\partial
n_F}{\partial E}\right)\nonumber\\
&&\times(\cos\theta;\sin\theta)
\frac{\nu^{eq}-\nu}{\omega_0\tau} .\label{Fchereznu}
\end{eqnarray}
Using the KE we can substitute the last fraction in the rhs by
$\partial\nu/\partial\theta$. This angle derivative of $\nu$ has sharp
peaks near the nodes, and simple integration shows that in the
very clean regime
\begin{equation}
d^\perp_{SF}=d^\parallel_{SF}=\frac{2}{\pi} .
\label{universal}
\end{equation}

\subsection{Extremely clean limit}

In the extremely clean limit $\tau$ is very large, namely,
$\epsilon_0\tau\gg 1$. It means that the period of a quasiclassical
trajectory $t_0$ is much less than $\tau$. Therefore, $\nu$ does not
have enough time to change its value during the motion, i.e.,
$\nu=\mbox{const}$. Since the solution of the KE (\ref{KEnu}) should be
an odd function of $\theta$, this constant is in fact $0$. So, to the
first order the total force is given by
\[
{\bf F}_{SF}=2\kappa C_0v_n\int \frac{dEd\theta}{2\pi}
\left(-\frac{\partial n_F}{\partial E}\right) 
\frac{\nu^{eq}}{\omega_0\tau} (\cos\theta;\sin\theta).
\]
At low $T\ll T_c\theta_H$
the leading contribution to the force in this
limit is dissipative:
\begin{equation}
d^\parallel_{SF}=\frac{1}{\epsilon_0\tau},
\qquad d^\perp_{SF}=\frac{\pi}{4(\epsilon_0\tau)^2} . \label{FEC}
\end{equation}
$\epsilon_0$ is the only characteristic of the spectrum, which enters the
expression for the force in this limit.

\subsection{Higher temperatures}

{\bf Moderately clean limit.} At higher temperatures $T>T_c\theta_H$
one should account for energy dependence of $\omega_0(\theta,E)$ at
characteristic $|E|\sim T$. In the moderately clean regime
$\omega_0^{max}\tau$ is still very small (this parameter depends on $T$
only through $\Delta_0$ and $\tau$) and the force is given by
Eq.(\ref{Ftotal}). To obtain corrections one should substitute
$\omega_0(\theta,0)$ in Eqs.(\ref{MCFdiss}),(\ref{MCdeltaFnd}) by
$\omega_0(\theta,E)$ and average over energies.\cite{EnergyAve} At low
$T\ll T_c$ these corrections remain of the same order of magnitude as
at $T\ll T_c\theta_H$. As $T$ approaches $T_c$ they decrease down to
$\omega_c\tau$ and $-(\omega_c\tau)^2$, respectively.

{\bf Superclean limit.} In the superclean regime
$\omega^{max}_0\tau\gg1$ the situation is more complicated. The time of
quasiclassical motion along the closed trajectory at a given energy
$E\gg\Delta'\theta_H$ has two contributions. The time of motion outside
$\theta_E$-vicinities of gap nodes is on the order of $2\pi/\epsilon_0$
(cf. Section 2), while the motion inside these vicinities takes time on
order of $8\theta_E/\omega_c\gg2\pi/\epsilon_0$.

{\it Low fields.} If $\epsilon_0\tau\ll1$ then the sample is in the
very clean limit (subsection B) at all temperatures, i.e., the system
(\ref{quas_motion}) spends a large amount of time (in units of $\tau$)
near the nodes and a small amount of time between the nodes. Therefore,
the solution of the KE is of the same form as in the subsection B.  In
$\theta_E$ vicinities of the nodes $\nu\equiv\nu^{eq}$.  Using the KE
(\ref{KEnu}) it is easy to show that on the interval
$[\theta_0+\theta_E; \theta_0+\pi/2-\theta_E]$ the solution is given by
$\nu=\nu^{eq}(\theta_0+\pi/2-\theta_E)$ to the accuracy $o(1)$. The
transition between these two types of behavior of the solution occurs
in an intermediate region of the width $\theta_\tau\ll1$.

There are two contributions to the integral over $\theta$ in
(\ref{Fchereznu}):
one from $\theta_E$-vicinities
of the nodes, and the other from the boundaries of these regions
where $\nu$ jumps to its equilibrium value in the internode region. The
integration shows that (to the accuracy $o(1)$)
\begin{eqnarray}
d^\parallel_{SF}&=&\frac{4}{\pi}\int_0^{\Delta_0}
dE\left(-\frac{\partial n_F}
{\partial E}\right)\left(1-\frac{E}{\Delta_0}\right) \label{ddfiniteT}\\
d^\perp_{SF}&=&\frac{4}{\pi}\int_0^{\Delta_0}
dE\left(-\frac{\partial n_F}
{\partial E}\right) \left(\arcsin\frac{E}{\Delta_0}+
\sqrt{1-\frac{E^2}{\Delta_0^2}} \right) \nonumber \\
&&+1-\tanh\frac{\Delta_0}{2T} .\label{drfiniteT}
\end{eqnarray}
The last line of Eq.(\ref{drfiniteT}) represents the contribution of energies
$|E|>\Delta_0$ and is exactly the same as in the s-wave case
(cf. Ref.\onlinecite{Parts}).

At very low temperatures $T\ll\Delta_0$ these expressions give 
(\ref{universal}). We see that the SF force is almost
$T$-independent in this region of temperatures. In the opposite limit
$T\gg\Delta_0(T)$ (which corresponds to $|T_c-T|\ll T_c$) we have
$d^\parallel_{SF}=0$ and $d^\perp_{SF}=1$ regardless of the value of
$\omega_0^{max}\tau$: in this region the SF force
exactly compensates for the Magnus force, and the vortex does not
experience any force in the limit of the normal state (note, however,
that our approach is not applicable to a very narrow vicinity of $T_c$
where $\Delta_0\tau\ll1$).

{\it Higher fields.}
In the other case $\epsilon_0\tau\gg1$ the time of periodic motion is
small at sufficiently low temperatures $T\ll T_c\omega_c\tau$. In this
limit the situation is similar to that of subsection C, and spectral
flow is suppressed. To obtain an expression for the force one should
average over energies\cite{EnergyAve}
Eq.(\ref{FEC}) with $\epsilon_0$ substituted by
$\epsilon_0(E)=\pi\omega_c/4\theta_E$. Evaluation shows that:
\begin{equation}
d_{SF}^\parallel\sim \frac{T}{T_c}\frac{1}{\omega_c\tau},
\qquad d^\perp_{SF}\sim\left(\frac{T}{T_c\omega_c\tau}\right)^2 .
\label{FEC_T}
\end{equation}
At low $T\ll T_c$ one can calculate prefactors in these expressions:
\[
d^\parallel_{SF}=
\frac{8\ln 2}{\pi}\frac{T}{\Delta'}\frac{1}{\omega_c\tau},\qquad
d^\perp_{SF}=
\frac{4\pi}{3}\left(\frac{T}{\Delta'}\frac{1}{\omega_c\tau}\right)^2 .
\]

If $\omega_c\tau\gg1$ the temperature in this extremely clean regime
can reach the values $T\sim T_c$. At these temperatures the
contribution of the states above the gap $\Delta_0$ to
$d_{SF}^\parallel,d_{SF}^\perp$ is of the same order as the
contribution of the states below the gap.  At $T\sim T_c$ expressions
(\ref{FEC_T}) correspond to conductivities of an s-wave superconductor
at $T\sim T_c$ or a normal metal in the limit $\omega_c\tau\gg1$.

If $\omega_c\tau\ll1\ll\epsilon_0\tau$ another regime is possible: at
higher temperatures $T_c\omega_c\tau\ll T<T_c$ the quasiclassical
motion near the nodes becomes slow (on $\tau$ time scale), the system
is in the very clean regime, and the force is described by
Eqs.(\ref{ddfiniteT}), (\ref{drfiniteT}).

\section{On the density matrix}

The parameter $\epsilon_0\tau$ is determined by the interlevel
distance, while the relation of $\Omega_0\tau$ to the structure of the
spectrum is unclear.  In this Section we consider the general KE for
the density matrix in terms of exact (not quasiclassical) eigenstates
of the hamiltonian.  We demonstrate a solution of this equation and
evaluate characteristic parameters determining the strength of the SF.
We discuss the origin of the parameters $\Omega_0\tau$ and
$\epsilon_0\tau$.

The KE for the density matrix $\hat\rho$ is
\begin{equation}
\frac{\partial\hat\rho}{\partial t}=
i[\hat\rho,\hat{\cal H}]-\frac{\hat\rho-\hat\rho^{eq}}{\tau}.
\end{equation}
In the stationary case
\[
i\rho_{mn}(E_n-E_m)=\frac{\rho_{mn}-\rho^{eq}_{mn}}{\tau}
\]
and therefore
\begin{equation}
\left(\frac{\hat\rho-\hat\rho^{eq}}{\tau}\right)_{mn}=
\rho^{eq}_{mn}\frac{i(E_n-E_m)}{1+i\tau(E_m-E_n)}.
\end{equation}
Here $n,m$ enumerate the eigenstates of $\hat{\cal H}$.
In two extreme cases we have:
\begin{eqnarray}
i(E_n-E_m)\rho^{eq}_{mn}, &\qquad&\hbox{\rm if}\quad |E_m-E_n|\tau\ll 1,\\
-\rho^{eq}_{mn}/\tau, &\qquad& \hbox{\rm if}\quad |E_m-E_n|\tau\gg 1.
\end{eqnarray}

The SF force on the vortex is
\begin{equation}
{\bf F}_{SF}=\frac{1}{2}\sum\limits_{m,n} {\bf P}_{nm}
\left(\frac{\hat\rho^{eq}-\hat\rho}{\tau}\right)_{mn} \label{Fdm}
\end{equation}
where ${\bf P}_{mn}$ are matrix elements of the operator of linear
momentum.

At low energies $E=-\omega_0(\theta)l$ and quantization leads to a
hermitian operator $\hat{\cal
H}=i\omega_0(\theta)\partial_\theta+i\partial_\theta\omega_0(\theta)/2$
as a hamiltonian. The eigenfunctions of $\hat{\cal H}$ are
\begin{equation}
f_N(\theta)=\sqrt{\frac{\epsilon_0}{\omega_0(\theta)}}
\exp\left(-iE_N\int^\theta\frac{d\varphi}{\omega_0(\varphi)}\right)
\end{equation}
(cf. Refs.\onlinecite{VolPoisson,KopVol} where a non-hermitian operator
$i\omega_0\partial_\theta$ was used), $E_N=N\epsilon_0$ are determined
by periodicity of the wave functions. Matrix elements of the momentum
operator ${\bf P}_{nm}$ [which also determines the density matrix
$\hat\rho^{eq}=n_F(\hat{\cal H}-{\bf v}_n\hat{\bf P})$] depend only on
the difference $N=m-n$. In terms of $P^+=P_x+iP_y$ the force is given
by
\begin{equation}
F^+_{SF}=\frac{v_n}{4} \sum\limits_N (P^+_N)^2
\frac{iN}{1+iN\epsilon_0\tau} \label{FN}
\end{equation}
at low $T\ll T_c\sqrt{H/H_{c2}}$
(one can split summations over $n$ and $N=m-n$ in (\ref{Fdm})
provided that $T\gg\Omega_0$). To evaluate\cite{ExactForce}
the force at higher
temperatures one should substitute $\epsilon_0$ by $\epsilon_0(E\sim
T)\sim \omega_cT_c/T$. Matrix elements of $P^+$ are 
\[
P^+_N=k_F\int\limits_0^{2\pi}\frac{dx}{2\pi} e^{-iNx} e^{i\theta(x)}
\]
where $\theta(x=-\epsilon_0t)$ is defined by Eq.(\ref{toftheta}), and are
non-zero only for $N=4k+1$. Analysis shows
that for small $N$ (such that $N\sim1$) one has
$P^+_N= 2\sqrt{2}(-1)^kk_F/\pi N$.
$P^+_N$ deviates from this expression at
$N\sim\Omega_0/\epsilon_0\sim\theta_H^{-1}$
(these $N$ are on the order of the maximal value of $d\theta/dx$)
where it slightly increases, and
$P^+_N$ rapidly decreases for $|N|\gg\theta_H^{-1}$.
So, this dependence determines two energy scales: the interlevel distance
$\epsilon_0$ and $\epsilon_0(\Omega_0/\epsilon_0)=\Omega_0$.
Analyzing Eq.(\ref{FN}) one can single out the same three regimes as
in the previous Section.

In the {\it moderately clean} regime one can neglect the second term in
the denominator in (\ref{FN}). Two regions of $N$ contribute to (only
reactive) force: the region of small $|N|$, which gives a contribution
$2/\pi$ to $d^\perp_{SF}$, and the region $N\sim\Omega_0/\epsilon_0$
contributing $1-2/\pi$ to $d^\perp_{SF}$.

In the {\it very clean} regime $\epsilon_0\tau\ll1\ll\Omega_0\tau$ the
second term in the denominator of Eq.(\ref{FN}) suppresses the
contribution of the region $N\sim\Omega_0/\epsilon_0$ as this term is
on the order of $\Omega_0\tau\gg1$ in this region. On the other hand,
$N\sim (\epsilon_0\tau)^{-1}$ contribute to the dissipative force in
this regime, the contribution to $d^\parallel_{SF}$ being $2/\pi$.
Finally, in the {\it extremely clean} regime $\epsilon_0\tau\gg1$ the
denominator in (\ref{FN}) suppresses all the terms, and to the first
order the force is ${\bf F}_{SF}={\bf P}^{eq}/\tau$. Calculation of the
equilibrium value of the linear momentum (at given ${\bf v}_n$) leads
to the result (\ref{FEC}).

In the s-wave case $P^+_N\ne 0$ only for $N=1$, and Eq.(\ref{FN}) gives
the same result as in Ref.\onlinecite{Stone}.

Thus, we found that the reactive force is due to transitions between
the levels separated by distances a) on the order of interlevel
distance, and b) on the order of maximal value of $\omega_0(\theta)$.
However at finite relaxation rate $1/\tau$ only the terms with energy
difference $N\epsilon_0<1/\tau$ in (\ref{FN}) are effective.  The
dissipative force, naturally, comes from transitions over c) the
distance $1/\tau$. Contributions b) and c) compete with each other: the
former is suppressed if $\Omega_0\tau\gg1$, and the latter in the
opposite limit.  In the extremely clean limit all the transitions are
suppressed.

\section{Summary and discussion}

Our analysis shows that like in s-superconductors, in d-materials the
SF along anomalous branches of the spectrum of fermions
bound to vortex cores produces a contribution to the force on a
moving vortex. In the regime of extreme SF (fast relaxation
in the core) the value of the non-dissipative force is the same as in
the s-wave case. This result could be anticipated: in fact,
consideration of the action for a superfluid shows
\cite{VolovikSpFlowOrig}
that the SF force is proportional to the number of the
states below the Fermi level ($C_0/m$) regardless of the pairing symmetry.

In the superclean regime coexistence of bound states with states in the
continuous spectrum leads to new features in vortex dynamics.  As the
relaxation time grows at constant $T$ and $H$ (the system gets
cleaner), the spectral flow contribution to the reactive force is
gradually suppressed.  The motion of a vortex with respect to the
normal component (crystall lattice) produces a perturbation of the
Bogolyubov-de Gennes hamiltonian ${\bf v\hat P}$ leading to transitions
between energy levels. Transitions between the levels $n$ and $m$ do
not lead to a net loss of momentum of quasiparticles if the time
derivative $E_n-E_m$  of the phase of an off-diagonal element of the
density matrix $\rho_{mn}$ is much larger than the relaxation rate
$1/\tau$. Such transitions effectively contribute to the force only for
$|E_n-E_m|\tau<1$.  In s-wave vortices the operator of linear momentum
leads to transitions only between the neighboring bound levels with
energy spacing $\omega_0$, and the parameter $\omega_0\tau$ determines
the total force on a vortex (cf.  Ref.\onlinecite{Stone}).  In the
d-wave case the perturbation of the hamiltonian effectively leads to
transitions between levels with energy differences in the range
$[\epsilon_0;\Omega_0]$. So, features of vortex dynamics depend on two
parameters, $\Omega_0\tau$ and $\epsilon_0\tau$. The reactive force is
completely suppressed only in the extremely clean limit.

The dissipative contribution is small in the limits of very large
(extremely clean limit) and very small (moderately clean limit) $\tau$
and it acquires its maximal value in the intermediate (very clean) region.
In this limit dissipative and reactive contributions
are equal and independent on $\tau$ and the details of the spectrum.

In connection with this discussion of different symmetries of the
pairing state let us mention that our calculations are not sensitive to
the sign of the order parameter. Only the anisotropy of the absolute
value of the gap counts.  So, the results are equally applicable to
d-wave and extremely anisotropic s-wave superconductors (with gap
nodes).

Let us discuss some conditions and restrictions we used in our
calculations. We have mentioned above that simple quasiclassical
description of Andreev bound states is valid in the clean limit
$\Delta(\theta)\tau\gg1$. In the moderately clean limit the main
contribution to the force is due to the states away from the nodes.
Therefore, the contribution of vicinities of the nodes is inessential,
and the condition $\Delta_0\tau\gg1$ is sufficient. In the superclean
limit the main contribution comes from vicinities of the gap nodes of
the width $\mbox{max}(\theta_\tau,\theta_E)$.  The condition
$\Delta(\theta_\tau)\tau\gg1$ can be translated to $E_F\tau\gg1$, which
is even weaker than $\Delta_0\tau\gg1$.

In our calculations we omitted the logarithmic factors [cf.
Eqs.(\ref{omegasmall}),(\ref{SpectrumLargeEn})]. In fact these factors
are unimportant in the moderately clean limit, where mostly the regions
away from the nodes contribute, and in these regions the logarithms are
absent (see Section 2). In the superclean limit, where the contribution
of the states close to the nodes is essential, the logarithmic factors
could only affect the boundaries between the different regimes. It
follows from the derivation in Sections 4,5 that they could not change
the results in the very clean regime. As for the extremely clean limit,
the logarithms enter the result only through $\epsilon_0$.

We studied 2D case so far. To account for the third dimension (with
${\bf H}\parallel \hat z$) we can integrate over all modes of motion
along the third axis $\hat z$.  $z$-dependence of the spectrum of the
bound states is slow (within an order of magnitude), therefore our
results would not change qualitatively in 3D.

Using our results one can investigate temperature, magnetic field and
purity dependences of transverse $\sigma_H$ and longitudinal $\sigma_O$
conductivities, which are given by (cf. (\ref{eq1}) and
Ref.\onlinecite{KopninNQC}):
\begin{eqnarray}
\sigma_O&=&\frac{|e|c}{mH}\rho d^\parallel_{SF}  \label{sigmaO}\\
\sigma_H&=&\frac{ec}{mH}\rho (1-d^\perp_{SF}) . \label{sigmaH}
\end{eqnarray}
In the moderately clean limit the behavior of dissipative and
non-dissipative forces is analogous to that in s-superconductors. In
the superclean limit the values of the Ohmic and Hall conductivities
and their relaxation time, magnetic field and temperature dependences
differ from those in the s-wave case (at the same $\Delta_0$).

In the moderately clean regime $d^\parallel_{SF}$ and $1-d^\perp_{SF}$
are small and almost field independent, $\sigma_H, \sigma_O\propto
H^{-1}$, $\sigma_H\ll\sigma_O$ and the Hall angle is very small. The
superconductor is always in this regime ($\Delta_0^2\tau/E_F\ll1$ or
$T_c-T\ll T_c$) just below $T_c$, and this region could extend down to
some finite or zero [if $\Delta_0^2(T=0)\tau(T=0)/E_F\ll1$] temperature.
In the latter case the superclean limit is unreachable.

In the very clean limit the conductivities are given by
Eqs.(\ref{sigmaO}),(\ref{sigmaH}) and
Eqs.(\ref{ddfiniteT}),(\ref{drfiniteT}). This regime is a subregime of
the superclean one with an additional condition that $\epsilon_0\tau\ll
\mbox{max}(1;\sqrt{H_{c2}/H}T/T_c)$. If in addition $T\ll T_c$ then
$\sigma_O$ and $\sigma_H$ are $T$ and $\tau$ independent and $\propto
H^{-1}$ [cf. Eq.(\ref{universal})]. Experimentally the superclean
regime in high-$T_c$ compounds was reached, e.g., in
Ref.\onlinecite{Exp}.

In higher fields
[$\epsilon_0\tau\gg\mbox{max}(1;\sqrt{H_{c2}/H}T/T_c)$; extremely clean
limit] $\sigma_O$ is small and the Hall angle is close to $\pi/2$. In
this regime $\sigma_O\propto H^{-2} \mbox{max}(T/T_c;\sqrt{H/H_{c2}})$.
In all the three regimes the conductivities depend on $T$ directly and
through $\Delta_0$ and $\tau$ (see Section 4). The exact view of the
phase diagram in $T$--$H$ plane is determined by parameters of a
sample, in particular, by temperature dependence of the relaxation time
$\tau$.

It would be interesting to investigate the influence of anisotropy of
the Fermi surface on vortex dynamics. In the moderately clean limit
anisotropy should not affect mutual friction parameters.
\cite{VolovikSpFlowOrig,VolPoisson} However in the superclean regime it
may be important.

In conclusion, we presented an analysis of the contribution of the
quasiparticles bound to vortex cores to vortex dynamics in the mixed
state of d-superconductors with lines of gap nodes.
We found three regimes of vortex dynamics. The behavior of Hall and
dissipative (Ohmic) conductivities in the moderately clean regime is
similar to that in s-wave superconductors, while in the superclean
regime new features were found.

As this article was in preparation, I learned about a recent work by
N.Kopnin and G.Volovik \cite{KVnew} devoted to the same subject.  We
note that our results for finite temperatures $T>T_c\sqrt{H/H_{c2}}$
differ from those of Ref.\onlinecite{KVnew}.

\acknowledgments

It is my pleasure to thank N.Kopnin, S.Simon, M.Stone, and G.Volovik
for very useful discussions of the issues considered in this article
and E.Fradkin for stimulating questions. I am very grateful to
G.Volovik communications with whom helped me to correct a mistake in
the consideration of the very clean limit.  The work was supported by
NSF through the grants DMR 89-20538COOP, DMR-94-24511, and
DMR-91-20000.

\end{multicols}


\begin{references}
\bibitem{SF-term} In this article we use the term ``spectral flow'' to
refer to the contribution of the quasiparticles bound to the vortex
core to vortex dynamics in all regimes though it can be described by
motion of quasiclassical energy levels only in the moderately clean
limit.
\bibitem{about_cont} Using the term ``continuous'' below we refer to
states with energies above the gap $\Delta(\theta)$.
\bibitem{t-not} For simplicity we use the same notation $t$ as in
Eq.(\ref{KE0}) since the real time variable does not appear below.
\bibitem{Exp-strict} It can be shown rigorously that in a vicinity of a
node of the width $a\ll1$ the solution of the KE can be approximated by
$A+Be^{-t/\tau}$ to the accuracy $O(a)$.
\bibitem{EnergyAve} By averaging over energies we mean calculation of
the integral $\int dE (-\partial_En_F)\dots$
\bibitem{ExactForce} To calculate the force exactly one should again
substitute $\epsilon_0$ by $\epsilon_0(E)$ in Eq.(\ref{FN}) and average
it over energies.
\end{references}
\end{document}